\documentclass[%
 reprint,
superscriptaddress,
 amsmath,amssymb,D_1
 aps,
]{revtex4-1}

\usepackage[all]{xy}
\usepackage{graphicx}
\usepackage{dcolumn}
\usepackage{bm}
\usepackage{mathrsfs}
\usepackage{amsmath}

\usepackage[usenames,dvipsnames]{color} 
\usepackage{amssymb}
\usepackage{slashed}
\usepackage{graphicx}
\usepackage{float}

\usepackage{subfigure} 
\usepackage{diagbox}

\usepackage{hyperref}
\usepackage[mathlines]{lineno}

\begin{document}

\preprint{APS/123-QED}

\title{Cosmological singlet diagnostics of neutrinophilic dark matter}
\author{Ottavia Balducci}
	\email{ottavia.balducci@physik.uni-muenchen.de}
	  \affiliation{Arnold Sommerfeld Center for Theoretical Physics, Theresienstra{\ss}e 37, 80333 M\"unchen\\}
	  
\author{Stefan Hofmann}
 \email{stefan.hofmann@physik.uni-muenchen.de}
  \affiliation{Arnold Sommerfeld Center for Theoretical Physics, Theresienstra{\ss}e 37, 80333 M\"unchen\\}

\author{Alexis Kassiteridis}
\email{a.kassiteridis@physik.uni-muenchen.de} \affiliation{Arnold
  Sommerfeld Center for Theoretical Physics, Theresienstra{\ss}e 37,
  80333 M\"unchen\\}

\date{\today}

\begin{abstract}
The standard model of particle physics is extended by adding a
purely neutrinophilic dark sector. 
It is shown that theories which accommodate standard model neutrinos 
as dark radiation are resurrected. 
Sterile neutrinos bridge the visible and the dark sector 
and can keep their mutual interactions effective even after the epoch
of big bang nucleosythesis. This prolonged contact between dark matter
and standard model neutrinos solves the well-known small-scale
structure problems of the cosmological standard model. 
The dark sector presented in this work satisfies all experimental and
observational constraints from particle physics and cosmology.

\begin{description}
\item[PACS numbers]
04.20.Dw, 04.62.+v, 04.70.-s, 11.10.-z
\item[DOI] 10.1103/PhysRevD.98.023003
\end{description}
\end{abstract}

\maketitle


\section{\label{sec:level1}Introduction}
Although much is known about the properties dark matter is required to
possess, its precise nature still eludes us.  Standard model
extensions include particle candidates that are classified as cold and
warm dark matter, e.g.~sterile neutrinos \cite{Dodelson:1993je}, both
of which allow to form cosmic large-scale structures in accordance
with observations, but at the same time give rise to challenges on
small scales known as the ``missing satellite", ``cusp vs core" and
``too big to fail" problems; for an overview of these issues see Ref.
\cite{Vogelsberger:2015gpr}.

Hierarchical structure formation in cold dark matter (CDM) requires,
among other prerequisites, a kinetic decoupling from some radiation
component after big bang nuclesynthesis (BBN)
\cite{Vogelsberger:2015gpr}; for a general analysis of CDM models with
late decoupling see Ref.  \cite{Bringmann:2016ilk}.  Almost ten years
ago an extremely interesting idea appeared, questioning whether such a
radiation component is required to belong to the dark sector extending
the standard model (SM), or whether it could belong to the particle
spectrum of the standard model. As a response to this question
so-called leptophilic models arose \cite{Fox:2008kb} that could
address the small-scale crisis. It soon became clear that neither
charged leptons \cite{Shoemaker:2013tda} nor photons
\cite{Bringmann:2016din} can play such a role due to the strong
constraints from observations of the cosmic microwave background (CMB)
radiation. The focus turned on neutrinophilic theories, where the
communication channel with the dark sector is kept effective for a
sufficient time duration by SM neutrinos \cite{Shoemaker:2013tda}.  It
was already known that some neutrinophilic theories involving vector
mediators were able to solve the small-scale problems
\cite{vandenAarssen:2012ag}.  However, the parameter space of these
attempts is severely constrained by weak decays of SM bosons, and, as
a matter of fact, enjoys no intersection \cite{Laha:2013xua} with the
parameter space required to address the small-scale crisis.  Parallel
to this development, neutrinophilic theories with a scalar mediator
were also ruled out; see
Refs. \cite{Pasquini:2015fjv,Heurtier:2016otg}. As stated in Ref.
\cite{Bringmann:2016din}, the radiation component that is responsible
for keeping cold dark matter in kinetic equilibrium for a sufficiently
long period, if existent, should belong to the dark sector extending
the standard model.

The scientific objective of this article is a proof of concept for a
purely phenomenological model, which resurrects the neutrinophilic
theories and provides clear solutions to the large- and small-scale
structure formation problems, while respecting all constraints from
experiments and consistency considerations, as well as the SM
symmetries. To the best of our knowledge, there is no model in the
literature that successfully addresses cosmic structure formation,
satisfies all these constraints and employs SM neutrinos as the last
scattering partners for CDM, while still providing ultraviolet
completeness, zero dark photon/photon mixing at one loop
\cite{Foot:2014uba}, which is strongly constrained
\cite{Bringmann:2016din}, and compatibility with big bang
nucleosythesis \cite{Heo:2015kra}.

This article is organized as follows.  In Sec. \ref{sec:level1} the
standard model is extended by a phenomenological dark sector.  Its
particle spectrum is justified and the relevant interactions are
introduced.  The following section is devoted to calculating cross
sections and decay rates that are pertinent for the cosmological
implications of this model.  In Sec. \ref{sec:par-constraints} we
consider the constraints from particle physics and cosmology and
consider how they restrict the parameter space. This allows us to
compute the most important cosmological observables, like relic
abundance, decoupling temperature and damping scales, in Sec.
\ref{sec:cosmological-obs}. By comparing these quantities with recent
experimental results we give the mass spectrum of the theory in
Sec. \ref{sec:spectrum}. We summarize the results and conclude in
Sec. \ref{sec:conclusion}.

\section{\label{sec:level1}Singlets in the dark sector}
In this section, the dark sector and its coupling to the visible
sector are described in detail.

 The proposed extension of the standard model
 {$\mathcal{L}_{\rm{ext}}$} consists of a dark sector
 {$\mathcal{L}_{\rm{ds}}$} and a neutrino bridge
 {$\mathcal{L}_{\rm{nb}}$}
\begin{equation}
\mathcal{L}_{\rm{ext}}=\mathcal{L}_{\rm{ds}}+\mathcal{L}_{\rm{nb}}\;{}.
\end{equation}
The dark sector contains a fermion $F$, a sterile neutrino $n$ and
three real scalars $S,\Phi{},X$. One notices that this field content
admits the same number of degrees of freedom (d.o.f.) as in the
original vector neutrinophilic theory \cite{vandenAarssen:2012ag},
which was ruled out. All of these new fields are SM singlets. A
summary of the new particles and their charges is presented in Table
\ref{table-1}.

\begin{table}
\begin{center}
\begin{tabular}{|c|c|c|c|c|c|}
  \hline{}
  \diagbox{\footnotesize{Symm.}}{\footnotesize{Fields}}&{$F$}&{$n$}&{$S$}&{$\Phi$}&{$X$}\\\hline{}
  \tiny{\mbox{{$\text{SU}\left(3\right)\times{}\text{SU}\left(2\right)\times{}\text{U}\left(1\right)$}}}
  & \footnotesize{{$\left({\bf{}1},{\bf{}1},0\right)$}} &
  \footnotesize{$\left({\bf{}1},{\bf{}1},0\right)$} &
  \footnotesize{$\left({\bf{}1},{\bf{}1},0\right)$}
  &\footnotesize{$\left({\bf{}1},{\bf{}1},0\right)$}
  &\footnotesize{$\left({\bf{}1},{\bf{}1},0\right)$}\\ \hline
               {$\mathbb{Z}_{2}$}&{$-$}&{$+$}&{$+$}&{$-$}&{$+$}\\\hline
\end{tabular}
\end{center}
\caption{The particles in the dark sector and their charges: they are
  total singlets regarding the SM gauge group and carry zero
  charges. A ``{$+$}'' in the third row, indicates a
  {$\mathbb{Z}_{2}$}-even symmetry, while a ``{$-$}'' a
  {$\mathbb{Z}_{2}$}-odd one.}\label{table-1}
\end{table}

We postulate that these Lagrangians accommodate the usual canonical kinetic terms
\begin{eqnarray}
  \label{ds1}
  \mathcal{L}_{{\rm{ds}}}&\supset&\tfrac{1}{2}\overline{F}\left({\rm{i}}\slashed{\partial}-m\right)F+\tfrac{1}{2}\overline{n}\left({\rm{i}}\slashed{\partial}-M_{R}\right)n\nonumber{}\\ &&
  -\tfrac{1}{2}S\left(\square{}+m_{S}^{2}\right)S-\tfrac{1}{2}\Phi{}\square{}\Phi{}\nonumber{}\\ &&-\tfrac{1}{2}X\left(\square{}+m_{X}^{2}\right)X\,{}.
\end{eqnarray}
The cosmologically relevant dimension-four interactions are given by
\begin{equation}
  \label{ds2}
  \mathcal{L}_{{\rm{ds}}}\supset\tfrac{1}{2}g_{S}\overline{F}SF+\tfrac{1}{2}\overline{g}_{S}\overline{n}Sn+\tfrac{1}{2}g_{X}\overline{F}XF-
  \mathcal{V}[S,\Phi]\, .
\end{equation}
Here we note that since the model is based on purely phenomenological
reasoning, couplings that do not appear are assumed to be weaker than
feeble ones. The last part of the dark sector is a scalar potential
$\mathcal{V}[S,\Phi]$ that enables a phase transition in the proposed
sector; hence
\begin{equation}
\mathcal{V}[S,\Phi]= \tfrac{\ell}{4}\left(\Phi^{2}-\overline{v}^{2}\right)^{2}+ \tfrac{x}{4}\Phi^{2}S^{2}\;{},
\end{equation}
where {$\overline{v}$} is a real parameter.\\ The communication
between the dark sector and the SM is achieved via the sterile
neutrino bridge
\begin{equation}
  \label{nb}
\mathcal{L}_{\rm nb}={\rm i} y\bar{L}\sigma_2H^* n + {\rm
  H.c.}\quad{}.
\end{equation}
All mass parameters are positive. Finally, the theory should contain a
Majorana mass $m_L$ for the SM neutrino to accommodate the final
observed neutrino mass, $m_\nu,$ after a hybrid-type seesaw (type
I+II); for a gauge-invariant implementation see
Ref. \cite{Dev:2017ouk}.\\

 We now justify the new particles and interactions appearing in
 {$\mathcal{L}_{\rm{ds}}$} and {$\mathcal{L}_{\rm{nb}}$}.

The dark matter is assumed to be a SM singlet represented by a
Majorana fermion $F$. We choose {$F$} to be a Majorana fermion,
instead of a Dirac one, just to have a more minimal model; but, in
principle, both Majorana and Dirac fermions admit similar properties
during the thermal evolution of the theory and thus would be equally
good candidates. The construction of the dark sector ensures stability
of the dark matter particles, since an accidental {$\mathbb{Z}_{2}$}
symmetry on the {$F$} field arises.

In order to solve the large- and small-scale problems of the
$\Lambda$CDM cosmology, we introduce Yukawa interactions between the
dark matter field and the SM singlet real scalars $S$ and $X$. The
interaction mediated through $S$ provides the self-interacting nature
of this theory that solves the small-scale problems, while the late
decays of the $X$ particles lead to the desired dark matter relic
density.

As already mentioned, a further SM-singlet sterile neutrino {$n$} is
postulated in Eqs. (\ref{ds1}), (\ref{ds2}), and (\ref{nb}). The
motivation behind it is the following: we want the dark sector to
couple solely to neutrinos at tree level, and not to charged
leptons. This poses a potential problem, because it violates the gauge
group of the left-handed fermions. The sterile neutrino mass could be
generated through some dark Higgs model with vacuum expectation value
$v_{\rm d}$, so that {$M_R \sim v_{\rm d}^2/M_{\tilde{F}}$}, where
$\tilde{F}$ is some heavier fermion field, such that before the
electroweak phase transition the fermionic current is
conserved. However, a precise derivation of the origin of this mass is
beyond the scope of this paper. The sterile neutrino {$n$} is
represented by a Majorana fermion and is coupled to the SM neutrinos
in the usual manifestly gauge-invariant way introduced in Ref.
\cite{Schechter:1980gr} with coupling {$y$} as shown in Eq.
(\ref{nb}).\\ Figure \ref{interaction-plots} shows the Yukawa
interactions between the most important new particles in the dark
sector.
\begin{figure}
\includegraphics[scale=1.2]{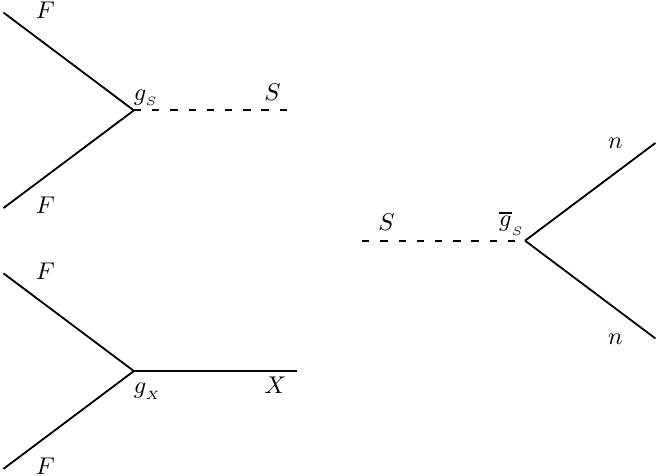}\label{interaction-plots}
\caption{The Yukawa couplings introduced in the extension to the
  Standard Model.}
\label{interaction-plots}
\end{figure}

The communication between the dark sector and the standard model is
then achieved through the gauge-invariant Yukawa portal given in
{$\mathcal{L}_{\rm{nb}}$}. For simplicity, we include in this work
only one lepton generation, as the generalization to all generations
is straightforward; moreover, we ignore the presence of the
Pontecorvo-Maki-Nakagawa-Sakata matrix. Let us stress already that the
sterile neutrino bridge constitutes the longest-lived communication
channel between the dark matter and the SM neutrinos. After
diagonalizing the neutrino mass matrix, we obtain the Majorana mass
eigenstates defined by the corresponding eigenstates of the $p^2$
Casimir. With some abuse of notation we call them $n$ and $\nu$ with
masses $M_R$ and $m_L - Y^2/2 M_R$ respectively, assuming that
$m_L,Y\ll M_R$, $ Y\equiv{}yv$ and $v\equiv \sqrt{2}\langle H\rangle$.

\begin{figure}
\includegraphics[scale=1.2]{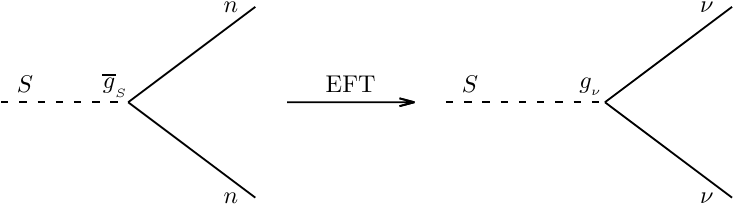}
\caption{After the electroweak phase transition the coupling
  {$g_{\nu}$} in the effective theory description (EFT) is obtained.}
\label{eft}
\end{figure}

In other words, the dark sector induces an effective coupling
$g_{_\nu}$ between the $S$ field and the SM neutrinos after the
electroweak phase transition
\begin{equation}
\tfrac{g_{\nu}}{2}\, S\, \overline{\nu}\nu
\end{equation}
with {$(g_{_\nu})^{1/2} = \left(\bar{g}_{_S}\right)^{1/2}\left(\delta
  r+ \mathcal{O}(\delta r^2) \right)$}, using the abbreviation
{$\delta r \equiv Y/\sqrt{2}M_R$}. This mechanism accommodates
naturally effective microcharges between the SM neutrinos and the $S$
bosons. Figure \ref{eft} shows the step from the coupling
{$\overline{g}_{S}$} to the effective coupling {$g_{\nu}$}. The above
effective description is valid at energies below the rest mass of the
sterile neutrino $n$, where this field can be integrated out. It is
this effective coupling $g_\nu$ that is responsible for the late
decoupling regime between the dark matter particles and the
SM neutrinos and for the correct cutoff masses of the protohalos
alleviating the ``missing satellite problem" and addressing positively,
together with the dark matter self-interactions, the ``cusp vs core"
and the ``too big to fail" issues.

\begin{figure}
\includegraphics[scale=1.2]{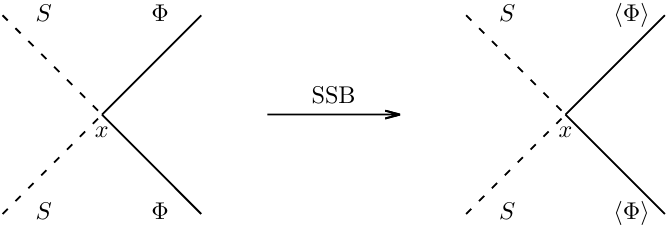}
\caption{The scalar field {$\Phi$} is introduced with a quartic
  coupling to {$S$} in order to change the mass of {$S$} dynamically
  by a spontaneous symmetry breaking with {$\left\langle{}\Phi\right\rangle{}=\overline{v}$}.}
\label{ssb-diagram}
\end{figure}

More precisely, in order to solve the small-scale structure problems
efficiently and simultaneously bypass all strong neutrinophilic
constraints \cite{Laha:2013xua,Pasquini:2015fjv,Heurtier:2016otg}, a
phase transition in the dark sector should take place, after the
protohalo formation. The most minimal realization is achieved after
adding an additional d.o.f.; nevertheless, more sophisticated
solutions (for example hidden global or local symmetries) could also
lead to the desired results, but since $\mathcal{V}[S,\Phi]$ works
well, we can keep it as it is. Therefore, we include in the spectrum
the real scalar field $\Phi$, which is solely coupled to the singlet
$S$ quadratically. Such a scalar potential enables a first-order phase
transition at times after big bang nucleosynthesis and could
potentially change the mass of the $S$ boson
significantly. Furthermore, we denote the temperature of this phase
transition by {$T_{c}$} and assume {$T_{c}\ll{}T_{\rm{EW}}$}, where
{$T_{\rm{EW}}$} is the temperature of the electroweak phase
transition. This is encapsulated in Fig. \ref{ssb-diagram}.  For
clarity, we denote with $a$ the scalar field after the phase
transition has taken place.

The terms of the effective Lagrangian after the symmetry breaking,
modulo self-interactions of $a$, that differ from the theory before
the breaking are
\begin{eqnarray}
\label{ssb}
\left.\mathcal{L}_{\rm{ds}}\right|_{\rm{SB}}&\supset&-\tfrac{1}{2}S\left(\square{}+M_{\text{obs}}^2\right)S-\tfrac{1}{2}a\left(\square{}+m_a^2\right)a\nonumber{}\\ &&-\tfrac{x}{4}a^{2}S^{2}-\tfrac{x}{2}\overline{v}aS^{2}.
    \end{eqnarray}
After the symmetry breaking, $a$ acquires a mass $m_a=\sqrt{2 \ell}
\bar{v}$. The singlet boson $S$ admits an observed mass of
$M_{\text{obs}} =\sqrt{m_{S}^2+x\bar{v}^2/2} $. This allows us to
write the critical temperature as {$T_{\rm c} \approx
  m_a^2/\sqrt{\ell}M_{\text{obs}} $} \cite{Carrington:1991hz},
assuming that $2m_{S}^2/x\bar{v}^2 \ll 1$. Note that both
contributions to {$M_{\text{obs}}$}, {$m_{S}^{2}$} and
{$x\overline{v}^{2}$}, are always positive. Figure \ref{table} gives a
summary of the newly introduced particles and the relations between
them.\\

Before concluding this section, we present the proposed mass
spectrum. We assume the following mass hierarchy in the dark sector:
\begin{equation} \label{massh}
 m_{X} \gg M_R> m,\,m_h \gg M_{\text{obs}}\gg m_{S}, m_a, m_\nu\;{}.
\end{equation}
For the purpose of this work, we decided to remain agnostic about the
origin of the initial boson and dark matter masses and, although
achievable, a self-consistent completion of the dark sector is
certainly beyond the scope of this paper. Such an arrangement allows
various decays, like for example {$n\rightarrow h\nu $}, {$S
  \rightarrow \nu \nu$}, {$X \rightarrow F F$}. Note that the relation
of {$m_a$} with the other masses appears as a natural consequence of
requiring a consistent thermal evolution of the theory. Furthermore,
the mass of {$S$} before the symmetry breaks, $m_{S}$, can be thought
of as a $t$-channel regulator in order to simplify the calculations.

\begin{widetext}
  \begin{center}
\begin{figure}
  \begin{center}
    \includegraphics[angle=90,scale=0.4]{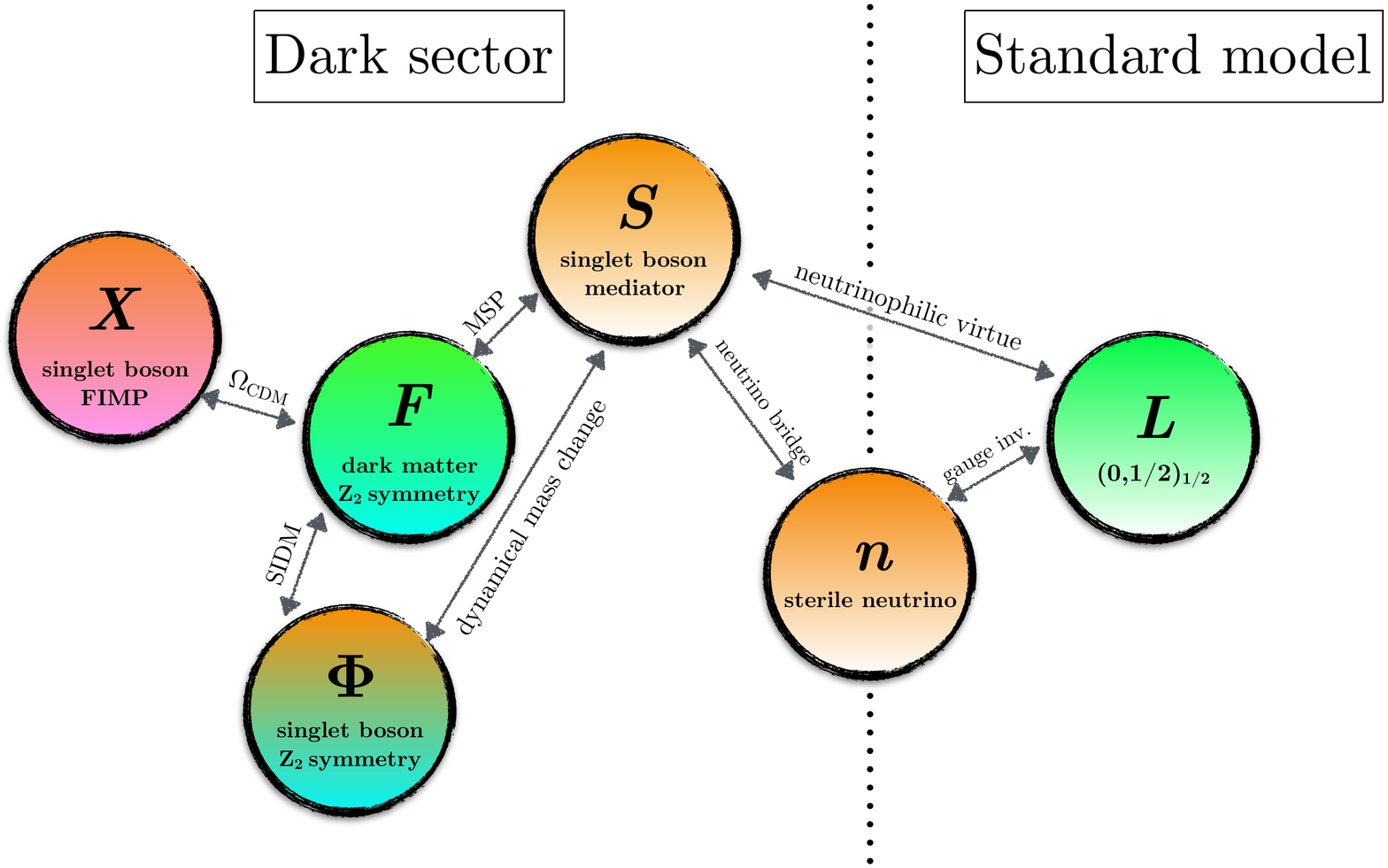}
    \end{center}
\caption{ A representation of the proposed
    extension to the standard model. The labels on the arrows explain
    the purpose of each (effective) interaction. Orange colored
    bubbles indicate that the particle decays, while green colored
    ones are stable. Particles with an accidental {$\mathbb{Z}_{2}$}
    symmetry are colored in light blue, while purple means that the
    particle was never in thermal equilibrium with the primordial
    plasma. }
\label{table}
\end{figure}
\end{center}
\end{widetext}

\section{Cross sections and decay rates}
This section is devoted to the presentation of cross sections and
decay rates relevant for particle physics and structure formation
processes in the primordial Universe, in which the dark sector
participates.

It is useful to introduce the dark sector Fermi constant of the theory
of neutrino scattering {$ G_D/\sqrt{2} \equiv g_{_S} g _{_\nu}/4
  m_{S}^2 $}. Annihilation into singlets $S$ implies a sink term in
the dark matter abundance that is characterized by the $p$-wave cross
section
\begin{equation}
\langle v_{\rm rel} \sigma_{\rm ann} \rangle = \frac{3}{8}\frac{\pi
  \alpha^2}{ m^2}\langle v_{\rm rel}^2 \rangle \quad{},
\end{equation}
where $\langle ... \rangle$ denotes the thermal average using relative
velocities, $\alpha = g_{_{S}}^2/4\pi$. In principle, if $m>M_R$ $F$'s
annihilate to $n$'s; however, this possibility is excluded by the
assumed mass hierarchy (\ref{massh}).

Finally, the dark matter abundance is created by $X$ decays with a
rate
\begin{eqnarray}
	\label{Phidecay}
	\Gamma_{X\rightarrow FF} = \frac{g_{_{X}}^2}{16\pi} m_{X} \; .
\end{eqnarray}	
This concludes the brief presentation of the dominant sinks and
sources for the dark matter abundance in the temperature regime $m\gg
T \gg m_{S}$, where $T$ is the average photon temperature.

 Elastic scatterings involving dark matter particles determine their
 kinetic properties and allow to quantify their deviation from local
 thermal equilibrium in terms of viscous processes sourcing entropy
 production. As mentioned above, the longest running channel in this
 respect is the one with SM neutrinos as scattering partners. At
 temperatures $m_{S}\gg T \gg m_\nu$, and at lowest order in the Fermi
 constant, the averaged momentum transfer elastic cross-section,
 defined by $\sigma_{ T}\equiv \int \mathrm{d}\Omega (1-\cos \theta)
 \mathrm{d}\sigma_{\rm el}/\mathrm{d}\Omega$, is given by
 \begin{eqnarray}
	  \langle v_{\rm rel} \sigma_T \rangle 
	 =
	 \frac{80 \zeta(5) }{ \pi \zeta(3)} G_D^2 T^2
	 \left(\frac{T_\nu}{T}\right)^2 
	 \; 
\end{eqnarray}
after applying the equipartition theorem. Note that in the above
expression all neutrino masses have been neglected. $T_\nu$ denotes
the neutrino temperature. The elastic scattering of {$S$} particles to
neutrinos is extremely suppressed. Indeed, although one can integrate
out the sterile neutrino field to obtain $g_{_\nu}$ for the physical
neutrinos, the internal two-point functions include sterile neutrino
propagators, which are not bare propagators but must be perturbatively
corrected with internal neutrinos. These rates are suppressed by a
factor of $g_{_\nu}^4$ for the IR-dominant part.

The numerical solutions in various regimes of the momentum transfer
cross section for fermions can be found in Ref. \cite{Tulin:2013teo}
and for the classical limit, $mv/m_{S} \gg 1$, we cite semianalytic
formulas for the corresponding total cross sections in different
kinematical regimes:
\begin{eqnarray}
  \sigma_{T}\approx{}\left\{
\begin{array}{lr}
  \frac{2\pi}{m_{S}^{2}}\beta^{2}\ln\left(1+\beta^{-2}\right)\,{},&\beta\lesssim{}1\\
  \frac{\pi}{m_{S}^{2}}\left(\ln\left(2\beta\right)-\ln\left(\ln\left(2\beta\right)\right)\right)^{2}\,{},&\beta\gtrsim{}1
\end{array}
  \right.
\end{eqnarray}
where $\beta\equiv 2\alpha m_{S}/(mv_{\rm rel}^{\; 2})$.  This
dependence seems to be of vital importance in order to resolve
small-scale anomalies that are typically present when structures form
in non-self-interacting dark matter (non-SIDM). Self-interactions
among the dark matter particles can be mediated via $X$ as well. In
this case, however, there is no significant Sommerfeld enhancement.

\section{\label{sec:par-constraints}Parameter constraints}
We proceed to the discussion of the constraints on the dark sector
arising from particle physics and cosmology.

\subsection{Constraints from particle physics}
The dark matter particles $F$ are assumed to be stable and produced in
local thermal equilibrium. Then partial-wave unitarity of the
scattering matrix bounds the annihilation cross section in the
primordial Universe, which in turn bounds the relic abundance and the
universal mass of the dark matter \cite{Griest:1989wd},
$m<\mathcal{O}(300)~$TeV. The effective number of neutrino generations
as measured by the Planck satellite implies a lower mass bound
$m>\mathcal{O}(1)~$MeV. Note that this bound holds precisely for the
dark matter introduced in the dark sector, since it is kept in local
thermal equilibrium for an extended period via elastic scatterings
with SM neutrinos. {\it A priori} there are no such bounds on the
sterile neutrino mass $M_R$ or on the masses $m_{S},m_{X}$, other than
the imposed mass hierarchy (\ref{massh}). Lower bounds on these mass
parameters do, however, arise in scenarios where the sterile neutrinos
and the dark bosons are kept in local thermal equilibrium with the SM
neutrinos after their respective kinetic decoupling from other SM
species. An extensive list of a variety of constraints on lepton-boson
interactions can be found in Ref. \cite{Shoemaker:2013tda}.

Let us turn to the particle physics constraints on the dark sector
couplings. The couplings {$g_{_{S}}$}, {$\overline{g}_{_{S}}$} and
{$g_{_{X}}$} are assumed to respect the perturbative domain. In
addition the coupling in the Yukawa bridge is bounded by $Y^2/2M_{R}
\sim m_L \lesssim 5$ GeV \cite{Dev:2017ouk}. Accordingly, in this work
we set the upper bound $\alpha\le 1/2$, which in turn implies an upper
bound on $m \approx 1.5~$TeV, as will be shown below. Note that the
dark sector does not facilitate a coupling between the dark bosons and
charged leptons. Therefore, $g_{_{\nu}}$ is only subject to
restrictions arising from the three-body decays of $Z^0$, $W^\pm$ and
from the $K^\pm$ decays, although such constraints are less tight than
those coming from cosmology. Furthermore, we anticipate that such
constraints should not be as strong as the light vector-neutrino
interaction bounds \cite{Laha:2013xua}, due to the absence of a
longitudinal contact term.

Consider the invisible three-body decay $Z^0\rightarrow\nu\nu S$ with
rate $ \Gamma_{Z^0\rightarrow\nu\nu S}\approx 0.18~{\rm GeV} \;
g_{_{{\nu }}}^{\; 2} (N_\nu/3)$ for an observed $S$-boson mass
parameter of $\mathcal{O}(10)~$MeV. The experimental error on $Z^0$
decays is $0.0023~$GeV \cite{Patrignani:2016xqp}. This allows to
constrain the effective neutrino coupling by {$g_{_{\nu}}\le
  0.12/\sqrt{N_\nu}$}. The result does not depend strongly on
$M_{\text{obs}}$. Note that if $S$ decays sufficiently fast, then the
four-body decay $Z^0\rightarrow 4\nu$ is dominant, which lightens the
previous constraint. The dark sector allows the decay $W\rightarrow\nu
e S$ as well. The experimental error on the decay of charged
electroweak gauge bosons is $0.042~$GeV \cite{Patrignani:2016xqp}.
For $M_{\text{obs}} \sim \mathcal{O}(10)$ MeV this implies $g_{_\nu}
\lesssim \mathcal{O}(1)$, which is not a significant constraint in our
context. Finally, the constraint on the effective neutrino coupling
arising from $K^\pm$ decays is more restrictive. In addition to the
standard decay {$K\rightarrow{}\nu\mu$}, the channel
{$K\rightarrow{}\nu\mu{}S$} is available as well. For masses {$m_{S}$}
of order {$\mathcal{O}\left(10\right)$} MeV and energies between
{$165.5$} and {$205.5$} MeV for the outgoing muon, the relative rate
is almost constant and can be explicitely written
\cite{Artamonov:2016wby} as
\begin{eqnarray}
	\frac{\Gamma_{K \rightarrow	\nu\mu S}}{\Gamma_{K \rightarrow  \nu\mu}}
	=
	7.4\times 10^{-4} g_{_{\nu }}^{\; 2} 
	\;.
\end{eqnarray}
The experimental bound on this ratio is $3.6\times 10^{-6}$, implying
a bound on the effective neutrino coupling $g_{_{\nu }}\le 7\times
10^{-2}$.

Note that, contrary to the situation analyzed in
Ref. \cite{Laha:2013xua}, there is no constraint from the elastic $\nu
e\rightarrow \nu e$ scattering, since the dark singlets in the dark
sector are decoupled from charged leptons at tree level. Additional
bounds on $g_{_\nu}$ due to lepton number violation and mesonic decays
are computed in Ref. \cite{Pasquini:2015fjv}. However, they are less
tight than the cosmological constraints, which will be discussed in
the next section.  Since {$g_{\nu}\lesssim{}0.12/\sqrt{N_{L}}$} and
{$\Gamma_{\text{el}}\sim{}N_{L}g_{\nu}^{2}$}, the extremal result for
{$\Gamma_{\text{el}}$} does not depend on the actual number of
neutrinos coupled effectively to the singlet boson; however, if $N_S$
denotes the number of identical $S$ bosons which are present, then
$g_\nu$ scales with $1/\sqrt{N_S}$.  Further bounds on
self-annihilation thermal averaged cross sections of $F$'s to
neutrinos can be found in Ref. \cite{Abbasi:2012ws}.

\subsection{Constraints from cosmology}
In this part, we study the constraints on the couplings $x,g_{_\nu}$
and $g_{_{X}}$, which arise from astrophysical observables.

The desired value of $M_{\text{obs}}$ lies in the sub-GeV/
multiple-MeV interval in order to solve the small-scale
issues. Furthermore, {$m_{S}$} should lie in the (sub-)keV range to
enable a late kinetic decoupling. Therefore, we discuss only the case,
where the phase transition in the dark sector takes place after the
kinetic decoupling of the $F$ particles from the neutrinos; otherwise,
the relevant interactions are highly suppressed. When the condensation
of {$\Phi$} happens after $T_{\rm kd}\sim \mathcal{O}(10^2)$ eV, the
$S$ particles annihilate rapidly to $a$ particles, as long as $m_a\ll
M_{\text{obs}}$. We demand that at the phase transition $S$ and $\Phi$
are in local thermal equilibrium. This condition implies the following
constraint for the coupling strength between {$\Phi$} and {$S$}:
\begin{equation}
x\gtrsim 10^{-13} \left(\frac{T_{\rm c} }{100\, \rm eV}\right).
\end{equation}
If the mass parameters {$m_a$} and {$m_{S}$} of  $a$ and the $S$
bosons respectively were below $\mathcal{O}(1)~$MeV, the presence of
these particles in the primordial plasma would modify big bang
nucleosynthesis. Furthermore, if these mass parameters were close to
the temperature of the primordial plasma at recombination, the cosmic
microwave background radiation would carry a fingerprint of the dark
sector. However, if the light particles of the dark sector and the
neutrinos freeze in after the end of big bang nucleosynthesis, at
around 30 keV, the impact is minimized. This leads to a conservative
upper bound for the effective neutrino coupling $g_{_\nu}^2 \lesssim
10^{-12}$ using the IR-dominant part of the elastic cross sections in
the absence of a fundamental cubic scalar interaction in the dark
sector. Furthermore, an important bound of {$g_{_\nu}\gtrsim 1.6
  \times 10^{-6} {\rm{MeV}}/M_{\text{obs}}$} arises for
$M_{\text{obs}}\sim\mathcal{O}(1)$ MeV, as computed from the
luminosity and deleptonization arguments regarding the observation of
SN1987A in \cite{Heurtier:2016otg}. In this model values of
$M_{\text{obs}} \approx 50$ MeV alleviate the cusp vs core and the
too big to fail problems together with a suitable set of
$\{m,g_{_X}\}$. Therefore, the final constraint on {$g_{\nu}$} reads
{$ 10^{-15} \lesssim g_{_\nu}^2 \lesssim 10^{-12}$}. Taking the
saturation limit of this relation, the effective neutrino coupling can
be eliminated as a free parameter; hence, we consider a typical value
of $g_{_\nu}^2= 10^{-13}$ in this work.

For simplicity we assume that the light dark sector particles are in
local thermal equilibrium after big bang nucleosynthesis and before
the phase transition takes place. After the condensation of {$\Phi$},
the dark $a$ emerges as a massive propagating d.o.f. The metastable
massive {$a$}'s contribute to the energy density of the Universe by a
factor of {$\Omega_a h^2 \approx m_a/300\, \rm eV$}. Nevertheless, the
condition that {$\Omega_{a}h^{2}$} be small enough is easily
fulfilled, since $m_a$ lies well below the eV scale.

Since the scalar field {$\Phi$} couples directly only to the $S$
boson, we turn our attention to the properties of the interactions
between $S,\Phi$ and the standard model. To test the above
construction, we calculate the deviation of the effective neutrino
d.o.f., which parametrizes the energy density of the
Universe, in different periods of its evolution. This is encapsulated
in the following definition, assuming that the neutrinos already
decoupled at $T_{\nu D} = 2.3$ MeV \cite{Enqvist:1991gx},
\begin{equation}\label{delta effective}
\Delta N_{\rm eff} \vert_{\rm BBN} = N_\nu \frac{\rho_{S+\Phi}}{\rho_\nu}\quad{}. 
\end{equation}
$\rho_i$ is the equilibrium energy density of the $i$th particle
species with initial conditions defined at $T_{\nu D}$ and including
all available d.o.f. This number is important because it parametrizes
the cosmic energy budget and can therefore be probed with high
precision. Let us introduce the abbreviation $\varepsilon\vert_{T_\nu
  D} \equiv \left(T_S/T_\nu\right)^{3} $.

Examining the parameter space we find that a common value of
$\varepsilon \sim 0.1$ is compatible with big bang nucleosynthesis
\cite{Cyburt:2015mya} and cosmic microwave background
\cite{Ade:2015xua} 1$\sigma$ measurements. Indeed one obtains $\Delta
N_{\rm eff} \vert_{\rm BBN} \approx 0.05$ and $\Delta N_{\rm eff}
\vert_{\rm CMB} \approx -0.02$ assuming that $T_{\rm c} > T_{\rm
  rec}$. This may also explain the recent tension about the decrease
of the deviation of effective neutrino number from BBN to CMB-based
measurements. A difference $\Delta N_{\rm eff}\vert_{\rm CMB}- \Delta
N_{\rm eff}\vert_{\rm BBN}<0$ is possible in the underlying
theory. However, if $T_{\rm c}< T_{\rm rec}$ and at the same time all
light d.o.f. are not in local thermal equilibrium with neutrinos at
the recombination period, then $\Delta N_{\rm eff}\vert_{\rm CMB}
\approx 0.05$; therefore, for optimal big bang nucleosynthesis/cosmic
microwave background compatibility the phase transition is should
happen before the recombination period with all light particles in
local thermal equilibrium. Moreover, we make the bound less tight by
demanding that $N_{\rm eff}+\Delta N_{\rm eff}\lesssim 3.5$
\cite{Fox:2008kb,Ade:2015xua}.  This sets an upper bound to the number
of singlet bosons, $N_S\leq 8$.
 
It is crucial that the $S$ boson should decouple from the primordial
plasma early enough, when all or almost all the d.o.f. of the standard
model are accessible. Quantitatively this means that the $S$-boson
decoupling from the SM plasma should take place before the QCD phase
transition, $T_{\rm freeze-out} > T_{\rm QCD}$, to obtain an allowed
value of $\varepsilon \lesssim 0.2$. This corresponds to a lower bound
for $M_R$ with respect to the couplings of the theory. We obtain up to
$\mathcal{O}(1)$ factors,
\begin{equation} \label{bound on M}
\left(\frac{g_{_\nu}}{10^{-6}}\right)\left(\frac{\bar{g}_{_S}}{1} \right)\left(\frac{ \rm TeV}{M_R}\right) \lesssim\mathcal{O}(1)
\; .
\end{equation}
In other words, light dark scalar bosons went through kinetic
decoupling from the SM species at temperatures $T_{S\, D}\gg
T_{\nu{D}}$ due to their IR-suppressed part of the cross section for
elastic scatterings with SM neutrinos. For instance, $T_{S\,
  {D}}=\mathcal{O}(10^6) T_{\nu{D}}$ for $M_R=\mathcal{O}(10)$ TeV,
$\bar{g}_{_S}=0.05$ and $g^2_{_{\nu }}\approx 10^{-13}$. In this
example kinetic decoupling from the SM species happens at
$T_{S{D}}\approx 0.5~$TeV.

The above discussion does not involve masses larger than $T_{\nu D}$
and as longs as $\tau_S \ll \tau_{\rm BBN}$ no particular constraints
arise. It turns out that the kinetic properties of the dark sector
depend on the mass of the light dark bosons relative to the neutrino
decoupling temperature, as will be worked out in great detail in the
next section. Anticipating distinct kinetic regimes, the dark sector
will be referred to as model $\mathcal{DS}_{\text{{\tiny{}M1}}}$ if
$m_{S}\le\mathcal{O}(1)~{\rm keV} \ll T_{\nu{D}}$, and as model
$\mathcal{DS}_{\text{{\tiny{}M2}}}$ if $m_{S}>T_{\nu{\rm D}}$. Let us
briefly turn to $\mathcal{DS}_{\text{{\tiny{}M2}}}$ and consider the
modification of the effective neutrino d.o.f. in this case,
\begin{eqnarray}
	\Delta N_{\rm eff}
	=
	\frac{60}{7\pi^4}
	\int\limits_{x_{S{\rm BBN}}}^\infty
	{\rm d}z\; z^2 
	\frac{\sqrt{z^2-x_{S{\rm BBN}}^{\; 2}}}{{\rm exp}(z)-1}
	\; ,
\end{eqnarray}
where $x_{S{\rm BBN}}=m_{S}/T_{\rm BBN}$. As an example, for
$m_{S}=10.5~$MeV we obtain $N_{\rm eff}=3.15$, which matches perfectly
with the value inferred from the Planck measurements
\cite{Ade:2015xua}. At the $2\sigma$ level, these also impose a lower
bound for the mass of the light dark bosons of $m_{S}>3.5~$MeV. In
other words, the light dark scalar bosons in
$\mathcal{DS}_{\text{{\tiny M2}}}$ are roughly 6 orders of magnitude
heavier compared to those in the $\mathcal{DS}_{\text{{\tiny M1}}}$
model. In the next section it will be shown how this renders
$\mathcal{DS}_{\text{{\tiny M2}}}$ less attractive from a
phenomenological point of view.

Finally, the feebly interacting massive boson $X$ should have a decay
channel into dark matter flavors after chemical decoupling at $x_{\rm
  f}$, and this channel should close before the epoch of big bang
nucleosynthesis. Using the corresponding decay rate (\ref{Phidecay}),
the following interval for the coupling of the massive scalar {$X$} to
the dark matter can be inferred:
\begin{widetext} 
\begin{equation}
\label{lambda constraint}
 4 \times 10^{-5} \left(\frac{T_{\nu D}}{2.3\, \rm MeV}
 \right)\left(\frac{g_*}{10.75}\right)^{1/4} \lesssim
 \left(\frac{g_{_{X}}}{3 \times 10^{-10}}\right) \left(\frac{m_{X}}{\rm
   PeV}\right)^{1/2} \lesssim \left(\frac{30}{x_{\rm f}}
 \right)\left(\frac{g_*}{108.75}\right)^{1/4}\left(\frac{m}{ \rm TeV}
 \right)\quad{},
 \end{equation}
\end{widetext}
where $g_*$ counts the effective d.o.f. in the primordial plasma. This
concludes the discussion of the coupling constraints.

\section{\label{sec:cosmological-obs}Cosmological observables}
In this section we calculate the relic abundance of the dark matter
candidates in the spectrum of the dark sector, their kinetic decoupling
temperature, both for $\mathcal{DS}_{\text{{\tiny M1}}}$ and
$\mathcal{DS}_{\text{{\tiny M2}}}$, and evaluate the characteristic
damping scales implied by kinetic decoupling and free streaming of the
dark matter candidates together with their cross sections for
self-interaction. Collisional and collision-free damping impact
structure formation and determine the properties of the typical first
protohalos. Finally, we investigate whether the dark sector
allows to resolve challenges for $\Lambda$CDM posed by the formation
of cosmic structure as it is observed. In this paper, we neglect a
possible second period of dark matter annihilation and the possible
formation of dark matter bound states
\cite{Bringmann:2016din}. Alternatively, one can consider out of
equilibrium dark matter production from late decays. We are only
interested in investigating the possibility of the existence of a
simple field theoretical model, which alleviates the large and small
structure problems. We note that all temperatures are given in the
photon plasma frame, unless stated differently.

\subsection{Dark matter relic abundance}
Let us first assume that the $F$ fields are the dominant dark matter
population, with annihilations into light singlet bosons as the
dominant sink for this population towards chemical
decoupling. ${\alpha}$ is fixed by requiring that the observed relic
density is correctly retrieved in the unbroken phase. The defining
equation of the dark matter distribution function $f(\textbf{p}(t))$
per d.o.f. of the underlying field in a
Friedmann-Robertson-Walker universe yields
\begin{equation}
(L-C)[f](\textbf{p})=0
\end{equation}
where $L:= p^0(\partial_t-H\textbf{p}\cdot \nabla_{\textbf{p}})$ is
the Liouville operator, which gives the change of $f$ with respect to
an affine parameter along a geodesic, and $C$ is the collision term.
We can simplify the Boltzmann equation by taking all the participating
particles up to the dark matter ones to admit equilibrium thermal
distributions. The Boltzmann equation for nonrelativistic chemical
decoupling leads to the change of the number density $n$ of dark
matter particles per entropy density $s$, denoted by $Y$, with respect
to the temperature $T\equiv m/x$ according to the Riccati-type
equation,
\begin{eqnarray}
	\label{Ric}
	\frac{{\rm d}Y_F}{{\rm d}x}
	=
	-\frac{E}{x^\ell}\left(Y_F^{\; 2} - {Y}_{F}^{{\rm eq}\; 2}\right)
	\; ,
\end{eqnarray}	 
where the efficiency is given by $E\equiv s(x)\langle v_{\rm
  rel}\sigma_{\rm ann}\rangle/H(x)\mid_{x=1}$, and $\ell=3$ for
$p$-wave annihilations, assuming approximately massless final states
[more precisely $m_{S}/(m/x)\ll 1$]. We define the chemical freeze-out value
$x_{\rm f}$ by demanding that
\begin{eqnarray}
	\langle v_{\rm rel}\sigma_{\rm ann}\rangle \; n_F^{\rm eq}(x_{\rm f})
	\stackrel{!}{=} H(x_{\rm f})
	\; .
\end{eqnarray}	
The exact solution of this equation is given in terms of the
Lambert-$W$ function, but a sufficiently good approximation can be
found iteratively. Already after one iteration:
\begin{eqnarray}
	\label{asol}
	x_{\rm f} &=& {\rm ln}\left(C\right) - \frac{\tfrac{1}{2}{\rm
            ln}\left({\rm ln}\left(C\right)\right)}
        {1+\tfrac{1}{2}{\rm ln}^{-1}\left(C\right)} \; , \nonumber
        \\ C &=& \sqrt{\frac{45}{32}}\frac{2M_{\rm Pl}\; m }
           {\sqrt{g_*}(T_{\rm f})\langle v_{\rm rel}\sigma_{\rm
               ann}\rangle^{-1}\mid_{x=1}} \; .
\end{eqnarray}	
This assumes that the number of relativistic d.o.f. varies slowly with
the temperature towards chemical decoupling. Inserting Eq.
(\ref{asol}) into Eq. (\ref{Ric}), the differential equation for $Y_F$
can be solved numerically for the coupling constant
$\alpha$. Requiring that $n_F(x_{\rm f})\gg n_F^{\text{eq}}(\infty)$,
and in the absence of feebly couplings, we obtain for the relic
abundance of the dark matter population as a function of $m$ and
$\alpha$,
\begin{eqnarray}
	\label{om}
	\Omega_F h^2 &\approx& {0.12}\;
        \left(\frac{\alpha}{0.1}\right)^{-2}\; \left(\frac{m}{{\rm
            TeV}} \right)^2 \; .
\end{eqnarray}	
The self-annihilation cross-section $\sigma_{\rm ann}$ is well below
the experimental sensitivity \cite{Abbasi:2012ws}. Note that Eq.
(\ref{om}) has been obtained assuming $g_{_{S}}\gg g_{_{\nu }}$ in
order to neglect annihilations into SM neutrinos. We also included the
Sommerfeld effect \cite{vandenAarssen:2012ag}, which gives corrections
of $\mathcal{O}(1)$ to the above result.

\begin{figure}
\includegraphics{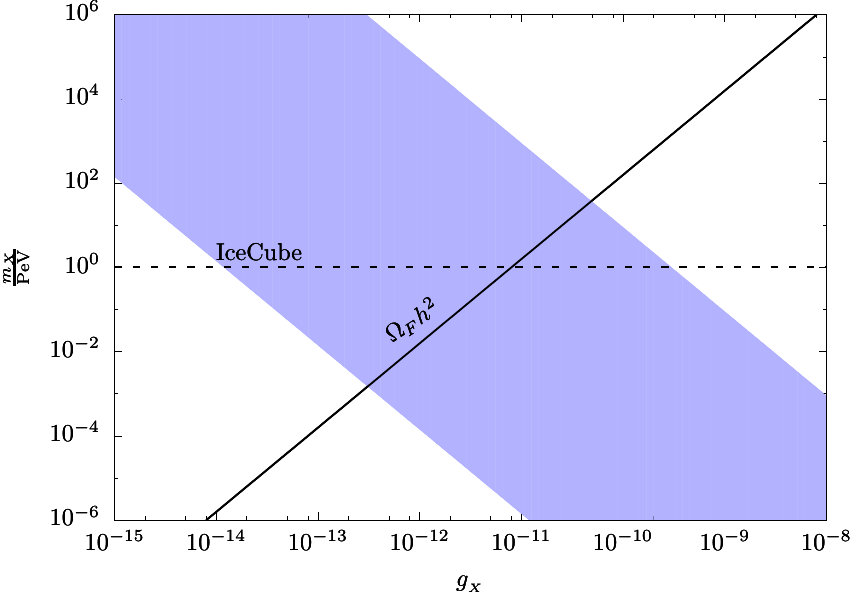}
\caption{Constraints on the relation between the coupling constant
  {$g_{_{X}}$} and the mass of the heavy scalar {$X$}. The shaded
  region represents the parameter space allowed by Eq. (\ref{lambda
    constraint}), while the solid curve comes from the condition for
  the correct relic density given in Eq. (\ref{relic-density}). The
  dashed line represents the recent IceCube measurements of
  ultra-energetic neutrinos in the {$\rm{PeV}$} regime. A dark matter
  mass of {$m=1\;{}\rm{TeV}$} is assumed.}
\label{constraints-gx}
\end{figure}

Let us stress that the constraints and the relic abundance (\ref{om})
fix $g_{_{S}}$ completely. However, in both models in order to resolve
the missing satellite problem, typically larger values of $\alpha$ are
needed for such masses; as a consequence, a different (primary or
subsidiary) mechanism to populate the dark matter is required, which
is still compatible with the framework above. The only option left in
the dark sector is to depart from the WIMP scenario as the exclusive dark matter production mechanism and consider the out of equilibrium decay of the heavy scalar
$X$. The number of $X$ particles per entropy varies with the
temperature $T=m_{X}/x$ as described by
\begin{eqnarray}
	\frac{{\rm d}Y_{X}}{{\rm d}x}
	\approx
	\frac{3}{8\pi^5 }\sqrt{\frac{5}{\pi}}
	\frac{M_{\rm Pl}\Gamma_{X\rightarrow FF}}{g_*^{\; 3/2}(T_{\rm f}) m_{X}^{\; 2}}
	x^3 K_1(x)
	\; ,
\end{eqnarray}	
where $K_1$ is the modified Bessel function of the second kind with
index one. The $\ell=-3$ behaviour is a characteristic property of
models with feebly interacting massive particles (FIMPs). From this we
find for the $F$ relic abundance,
\begin{eqnarray}
	\Omega_F h^2
	&\approx&
	{0.12}\left(\frac{g_{_{X}}}{8\times 10^{-11}}\right)^2
	\left(\frac{114}{g_*(T_{\rm f})}\right)^{3/2}
	\nonumber \\ 
	&&\times\left( \frac{m}{{\rm TeV}} \right)
	\left(\frac{{\rm PeV}}{m_{X}} \right)
	\; .\label{relic-density}
\end{eqnarray}	
It is very promising that this scenario, for values of
{$g_{{_{X}}}\gtrsim{}10^{-9}$}, is perfectly compatible with the
constraints on the $g_{_{X}}$ coupling (\ref{lambda constraint}), as
can be seen in Fig. \ref{constraints-gx}. Concluding, in the presence
of multiple heavy bosons, smaller couplings are needed, which allow
smaller dark matter masses for higher values of $m_{X}$ due to
Eq. (\ref{lambda constraint}). The same can be achieved after using
Dirac fermions for $F$ and/or upgrading $X$ to a vector boson.

\subsection{Decoupling temperatures}
As calculated above, chemical decoupling of the dark particles happens
long before the SM neutrinos decouple. However, this is not the case
for the cease of efficient momentum exchange: the kinetic coupling of
dark matter to neutrinos is active much longer. In order to have a
phenomenologically viable scenario, kinetic decoupling should be
scheduled for times long after big bang nucleosynthesis, but still
before the $\sim 100~$eV epoch, due to the constraints from the
Lyman-$\alpha$ measurements \cite{Baur:2015jsy}. We use simply the
constraints from the general overview \cite{Bringmann:2016ilk} and not
the most recent ones \cite{Irsic:2017ixq}, since such constraints are
described as overly restrictive; see Ref. \cite{Garzilli:2015iwa}.

A rough but sufficiently accurate estimate for the temperature $T_{\rm
  ls}$ of the typical last elastic contact between the dark matter and
the SM neutrinos, which represents the longest active communication
channel, is obtained by equating the corresponding elastic scattering
rate $\Gamma_{\rm el}\equiv \langle v_{\rm rel}\sigma_{\rm el}\rangle
n_\nu^{\rm eq}$ with the Hubble expansion rate. For example, assuming
three generations of SM neutrinos and $m_{S}<\mathcal{O}(1)~$keV, we
find $T_{\rm ls}\approx T_{\rm eq}$, where $T_{\rm eq}$ denotes the
temperature of matter-radiation equality. This characterizes the
typical spatial hypersurface from which dark matter enters the
free-streaming regime.
 
$T_{\rm kd}$ is the temperature at which the dark sector kinetically
decouples from the standard model, when the elastic interactions
between the dark matter particles and the neutrino generations cease
to sustain local thermal equilibrium \cite{Bringmann:2006mu}.  In
model $\mathcal{DS}_{\text{{\tiny M1}}}$ it depends on the parameters
$\{g_{_X}, m_{S}, m, g_{_\nu}\}$. Hence, the temperature at kinetic
decoupling is not completely fixed by the relic dark matter
abundance. Moreover, these parameters allow to accommodate other
constraints, like the one on {$\langle\sigma_T/m\rangle$}, to address
the cusp vs core problem.

We note that $T_{\rm kd}$ is not very sensible to small changes of
$m_{S} < \mathcal{O}(1)$ keV, but it changes rapidly with respect to
deviations of the neutrino effective coupling. Numerically, using the
rate of the averaged momentum transfer elastic cross section as in
Ref. \cite{Shoemaker:2013tda} and equating it to the Hubble rate, we
obtain $T_{\rm kd} \lesssim 0.6$ keV and at the same time the variable
set $\{g_{_{S}}, m, m_{S}, g_{_\nu}\}$ fulfills all previous
constraints. If a larger value of $\alpha$ is needed, then the FIMP
scenario comes into play and cures this issue. An elegant
instantaneous kinetic decoupling can be achieved by setting the
critical temperature equal to the desired $T_{\rm kd}$.

\subsection{Damping Scales}
Eventually, around a temperature $T_{\rm kd}$, elastic scattering
processes between dark matter and SM neutrinos happen too infrequently
to sustain local thermal equilibrium. Since the neutrino abundance is
much larger than the dark matter one, every nonrelativistic dark
matter particle will interact a huge amount of times with the
relativistic neutrinos until the time of kinetic decoupling. On the
other hand, hardly any neutrinos will ever interact with a dark matter
particle. Around {$10^{10}$} interactions between dark matter and
neutrinos are needed for a {$\mathcal{O}\left(1\right)$} momentum
transfer. This is easily achieved since the abundances of the two
types of particles are so different from each other and since there is
a long enough period of time until kinetic decoupling
\cite{Gondolo:2016mrz}. For this reason, any structure smaller than
     {$M_{\text{d}}=(4\pi/3) \rho_{\rm m}(T_{\rm kd})/H^3(T_{\rm
         kd})$} will be destroyed before {$T_{\rm kd}$}
     \cite{Loeb:2005pm}. We thus find the following estimate of the
     characteristic damping mass:
\begin{eqnarray}
	M_{\rm d}
	\approx
	2.7\times 10^8 \; 
	\frac{g_{*\;{\rm s}}(T_{\rm kd})}{g_*^{3/2}(T_{\rm kd})}\;
	\left(\frac{\rm keV}{T_{\rm kd}}\right)^3 \; M_\odot
	\; ,
\end{eqnarray}	
where $g_{*\;{\rm s}}(T)$ and $g_* (T)$ denote the effective number of
relativistic d.o.f. contributing to the entropy and energy density at
temperature $T$ respectively.

A relevant independent quantity is the free-streaming length of dark
matter particles \cite{Green:2005fa}. However, we expect that the
acoustic damping is much more efficient than free streaming for very
late decoupling temperatures, since the latter has a $1/T^{3/2}_{\rm
  kd}$ dependence in contrast to the $1/T^{3}_{\rm kd}$ dependence of
$M_{\rm d}$.  The mass of the smallest possible protohalo, that could
be formed, is found by taking the largest of the above two masses. A
possible solution to small-scale abundance problems of $\Lambda$CDM
cosmology, i.e. the missing satellite problem, can be found after
suppressing the power spectrum at scales as large as that of dwarf
galaxies \cite{Aarssen:2012fx}, which is provided by damping masses of
order {$\log_{10}\left[ M_{\rm d}/M_\odot \right] \gtrsim 9$} and not
larger than $\sim 10^{10} M_\odot $ as stated in Refs.
\cite{Bringmann:2016ilk,Baur:2015jsy,Bullock:2010uy,Shoemaker:2013tda,Aarssen:2012fx}. For
a thorough discussion see Ref. \cite{Piceno:2015jao}. It should be
clear that if a stronger upper bound is activated, then it is much
easier to find a common parameter set to achieve such kinetic
decoupling temperatures; nevertheless, we do not examine such a case
in this work.

In model $\mathcal{DS}_{\text{{\tiny M1}}}$, the mass scale of
acoustic damping \cite{Hofmann:2001bi} can be easily fixed at $10^{10}
M_\odot$.  In model $\mathcal{DS}_{\text{{\tiny M2}}}$, the cut-off
masses are much smaller, namely we approximately obtain $ M_{\rm d\,
  \mathcal{DS}_{\text{{\tiny M2}}}}\approx 5\times 10^{3}
M_\odot$. These values are far from promising. Although they allow
protohalo masses that are not excluded by current collider and direct
search constraints, they do not provide a solution to the missing
satellite problem. We note that for neutralinos in the minimal
supersymmetric SM the predicted masses are around the Earth's mass
\cite{Green:2005fa}. Such masses are well below the masses that the
current numerical simulations studying the large-scale structures can
resolve \cite{Gondolo:2016mrz}.

Concluding, we notice that the model $\mathcal{DS}_{\text{{\tiny
      M2}}}$ is ruled out by astrophysics, if one assumes that the
astrophysical constraint of order 9 is actually the lower bound. This
does not alleviate the enduring missing satellite problem respecting
the Planck measurements as stated in Ref. \cite{Aarssen:2012fx}; in
other words, at least $m_{S} < \mathcal{O}(1)$ keV is needed in order
to generate masses near $ 10^{10} M_\odot$, which leads us to model
$\mathcal{DS}_{\text{{\tiny M1}}}$. The FIMP scenario in
$\mathcal{DS}_{\text{{\tiny M1}}}$ makes almost the whole parameter
set $\left\{g_{_{S}},m\right\}$ accessible, and hence one can always
obtain $M_{\rm d\, \mathcal{DS}_{\text{{\tiny M1}}}} \sim
10^{10}M_\odot$. On the contrary, a dominant WIMP production mechanism
fails to deliver the necessary values of {$g_{_{S}}$}. Therefore, the
missing satellite problem is easily solved for the FIMP case for
sub-keV mediators, by considering for example {$m=300\,{}\text{GeV}$},
{$m_{S}=0.1\,{}\text{keV}$} and {$\alpha\approx{}0.1$} or
{$m=1\,{}\text{TeV}$}, {$m_{S}=10\,{}\text{eV}$} and
{$\alpha{}\approx{}1/\pi$}. All effective neutrino couplings are in
perfect accordance with Ref. \cite{Heurtier:2016otg}. Alternatively,
tuning $T_{\rm c}$, we assure that an instantaneous late kinetic
decoupling at the desired temperature $T_{\rm c} \approx T_{\rm kd}$
takes place.

\subsection{Effects of SIDM cross sections on dwarf galaxies}
The final test of the dark matter models is given here: we investigate
whether this parameter set supports noticeable effects on dwarf
galaxies. As we mentioned before, the SIDM elastic cross section in
galaxy clusters should lie within a constrained range of
values. Quantitatively, the cross section for elastic scatterings
between the dark matter particles should lie in the interval
{$\langle\sigma_T/m\rangle_{v_{\rm therm}}\in
  [0.1,10]~\text{cm}^2\text{g}^{-1}$} \cite{Bringmann:2016din} to
resolve the cusp vs core problem as shown in
Ref. \cite{Zavala:2012us}, where the indicated average is taken with
respect to a Maxwell-Boltzmann distribution with $v_{\rm
  therm}=\mathcal{O}(10^{-4})$ as the most probable velocity (or
around $10^{-2}$ for massive clusters). For a detailed analysis see
Ref.  \cite{Tulin:2017ara}.

Exploring the parameter space of this family of models, we find that
the values of SIDM elastic cross sections in model
$\mathcal{DS}_{\text{{\tiny M2}}}$ are of the desired order, $ \sim
0.5-1 \, {\rm cm}^2 {\rm g}^{-1}$ at dwarf scales. Therefore, after
setting $x, g_{_{X}}$ and $\ell$ to zero without loss of generality,
no further modification of the dark sector is needed. However, sub-keV
mediators need either $\mu$-dark matter charges or multi-TeV values of
$m$ \cite{Tulin:2013teo}. Both possibilities are strictly excluded, as
found previously, due to the previous constraints, thermal evolution
or perturbation theory. Therefore, in model
$\mathcal{DS}_{\text{{\tiny M1}}}$, as we mentioned before, we make
use of the scalar potential, which should admit a phase transition at
$T_{\rm c}\lesssim T_{\rm kd}$. This does not have an effect on the
damping masses, but cures the SIDM values automatically. For the
relevant subspace of parameters, the obtained values can stay always
inside the constrained interval as one sees in
Ref. \cite{Tulin:2013teo}. The SIDM cross-sections between $ 0.1-10 \,
{\rm cm}^2 {\rm g}^{-1}$ are easily accessible for values of DM
resonances in the sub-TeV/TeV range together with sub-GeV
mediators. These results are in line with the maximum circular
velocities observed and the values affecting dwarf galaxy scales
\cite{Bringmann:2016din}. For example, for the previous parameter set
considering the FIMP scenarios with {$m=0.3(1)\,{}\text{TeV}$}, a
value of $M_{\text{obs}} = 5(3)$ MeV leads to the thermal averaged
SIDM cross section values {$ \gtrsim 1.0 \, {\rm cm}^2 {\rm g}^{-1}$}
at the dwarf scales and {$ \lesssim 0.1 \, {\rm cm}^2 {\rm g}^{-1}$}
at cluster scales providing at the same time potential solutions to
all enduring small structure problems of the $\Lambda$CDM model, as
stated in Ref. \cite{Tulin:2017ara}. We stress that most of the
parameter space from Fig. 3 in Ref. \cite{Kaplinghat:2015aga} is
accessible in {$\mathcal{DS}_{\text{{\tiny M1}}}$} (FIMP scenario);
for example, {$m\approx{}20\,{}$}GeV,
{$\alpha\approx{}\alpha_{\text{em}}$} and {$m_{S}\approx{}10\,{}$}keV
lie inside the desired SIDM interval and provide
{$M_{\text{d}}\approx{}10^{10}M_{\odot}$}.\\ Finally, we note that
after choosing the parameters to be {$m\approx{}1\;{}\rm{TeV}$},
{$\alpha\approx{}1/\pi$}, {$m_{S}\approx{}28\;{}\rm{eV}$} and
{$M_{\rm{obs}}\approx{}3\;{}\rm{MeV}$}, leading to
{$T_{\rm{kd}}\approx{}420\;{}\rm{eV}$}, we are able to reproduce the
ETHOS-4 model \cite{Vogelsberger:2015gpr}, which is cosmologically
compatible and solves the missing satellite issue, the too big to fail
problem and the cusp vs core problem.

\section{\label{sec:spectrum}The underlying spectrum}
In the present section we examine the resulting spectrum of the
theory, which is able to provide solutions to the $\Lambda$CDM
problems.

Besides the stable $F$, both the bosons and the sterile neutrino
decay. In addition, for perturbative couplings {$ \bar{g}_{_{S}}/5=y
  =1/10$} and $g_{_\nu}^2=10^{-13}$, the sterile neutrino admits a
mass $M_R \approx 20$ TeV and $\delta r \approx 10^{-3}$. For this
parameter set one finds that the $S$-singlet decouples from the
SM-neutrino plasma at very high temperatures, around $ m,M_R$ , due to
the tiny value of the effective neutrino coupling $g_{_\nu}$; at that
time almost all the d.o.f. of the standard model are
relativistic. This means that {$S$}, {$\Phi$} and {$X$} freeze out
early enough and there is not much time left for them to achieve local
thermal equilibrium with neutrinos, which relaxes the strong
constraints on $M_{\text{obs}}$ from Planck measurements
\cite{Heo:2015kra}. This is a unique feature of this family of models,
since the $S$ interactions at tree level include two-point functions of
$F$'s or $n$'s, which at the zero-momentum limit are severely
suppressed due to the large masses, $m$ and $M_R$. Furthermore, the
mass of {$a$} is of order $m_a \sim \mu$eV.

Recently, the IceCube Collaboration published results about
extraterrestrial ultra-energetic neutrinos, i.e. Bert, Ernie
\cite{Aartsen:2013bka} and Big Bird with approximate energies in the
PeV regime. We could easily accommodate these observations in
$\mathcal{DS}_{\text{{\tiny M1}}}$: $X$ admits a mass of $m_{X} = 1$
PeV together with a minimum DM mass of $m=300$ GeV; additional FIMP
sources, i.e., more than one heavy $X$ boson, allow lower values of
$m$ due to the constraints on the $g_{_{X}}$ coupling (\ref{lambda
  constraint}).  We stress that this parameter set still resolves all
three sub-scale structure problems simultaneously after $T_{\rm c}$,
delivering damping masses and SIDM cross sections of the desired
order, together with the predicted value of $\Omega_{\text{CDM}}$.

\section{\label{sec:conclusion}Conclusion}
In this work we presented a proof of existence of a purely
phenomenological model, which resurrects the neutrinophilic theory
solving the dark matter large- and small-scale problems, while
respecting the experimental bounds and the standard model
symmetries. To the best of our knowledge, there is no model which uses
SM neutrinos as the last scattering partner for CDM. The proposed
model admits only dimension-four operators, respects all cosmological and
particle physics constraints and alleviates the small-structure
problems of the $\Lambda$CDM cosmology.

We introduced a singlet family of purely neutrinophilic dark matter
models using SM neutrinos as dark radiation, thus proving that
neutrinophilic theories are not over. The proposed dark sector
interactions work as an extension of the SM of particle physics, while
respecting the SM symmetries. We computed astrophysically important
observables (cross sections and decay rates) in this framework and we
presented and explained a list of constraints on the parameter space
emerging from the particle physics and cosmological nature of the
models. Finally, we estimated the chemical and kinetic decoupling and
last scattering temperatures and we determined the relevant smallest
protohalo masses. These masses are far above the Earth's mass produced
by the usual WIMP models. The derived $M_{\rm d}$ alleviates the
``missing satellite'' problem and together with the SIDM cross
sections provides a clear solution to the cuspy profile and massive
subhalos issues of {$\Lambda$}CDM simultaneously.

In particular, the sterile neutrino works as the gauge-invariant
bridge between the standard model and the dark sector.  It couples the
SM-neutrino species to the stable dark matter candidate, $F$, with a
strength compatible with particle physics constraints.
Out-of-equilibrium decays of the heavy $X$-bosons achieve the required
relic abundance of dark matter.  The ``missing satellite" problem is
alleviated by the light singlet $S$ and the late decoupling properties
of the SM neutrinos.  The ``cusp vs core" and the ``too big to fail"
issues are solved after the condensation of the scalar $\Phi$ happens.

It is worth noticing that in this model the strongly constrained
SU$(2)_L$ violating $\nu-S$ coupling arises naturally as an effective
interaction. Furthermore, the case of a simple dominant WIMP
production mechanism does not solve all the problems simultanesouly
and is ruled out, as we showed, at least in the perturbative regime of
the coupling constant $\alpha$.  Nevertheless, the FIMP scenario is
able to alleviate all small-scale issues of the CDM paradigm. In
addition, both dark matter production schemes provide excellent big
bang nucleosynthesis/cosmic microwave background compatibility.
Therefore, $\Lambda$CDM models are still viable and seem more
attractive in this neutrinophilic perspective.

The straightforwardness of this phenomenological model is
intriguing. In future works we hope to return to this model,
investigate the possibility of a more minimal construction and
consider different spectrum hierarchies.

\begin{acknowledgments}
It is a great pleasure to thank Frederik Lauf and Kerstin Paech for
inspiring discussions. We are grateful for comments by Torsten
Bringmann who helped to clarify the presentation of our work. We
appreciate financial support of our work from the Deutsche
Forschungsgemeinschaft (DFG) cluster of excellence ``Origin and
Structure of the Universe'', the Humboldt Foundation, and from
Transregio 33 (TRR 33) ``The Dark Universe''. O.B. is grateful for
financial support from the ``FAZIT-STIFTUNG''.

\end{acknowledgments}

\end{document}